# Room-temperature coherent optical manipulation of single-hole spins in solution-grown perovskite quantum dots

Xuyang Lin[1,2,†], Yaoyao Han[1,2,†], Jingyi Zhu[1], Kaifeng Wu[1,2]*.

[1] State Key Laboratory of Molecular Reaction Dynamics, Dalian Institute of Chemical Physics, Chinese Academy of Sciences, Dalian, Liaoning 116023, China.

[2] University of Chinese Academy of Sciences, Beijing 100049, China.

*Correspondence to: kwu@dicp.ac.cn

**Manipulation of solid-state spin coherence is an important paradigm for quantum information processing. Current systems either operate at very low temperatures or are difficult to scale-up. Developing low-cost, scalable materials whose spins can be coherently manipulated at room temperature is thus highly-attractive for a sustainable future of quantum information science. Here we report ambient-condition all-optical initialization, manipulation and readout of single-hole spins in an ensemble of solution-grown $CsPbBr_3$ perovskite QDs. Single-hole spins are obtained by sub-picosecond electron scavenging following a circularly-polarized femtosecond-pulse excitation. A transversal magnetic field induces spin precession, and a second off-resonance femtosecond-pulse coherently rotates hole spins via strong light-matter interaction. These operations accomplish nearly complete quantum-state control of single-hole spins at room temperature.**

Coherent control of spins in solid-state systems holds great promise for quantum information science (QIS).[1] Compared to bulk semiconductor materials, localized systems, such as epitaxial-grown quantum dots (QDs), were considered to be more adaptable to QIS because of the possibility to address and manipulate single spins.[2-4] Additional examples of localized systems include defect centers or dopants in solids.[5] Traditionally, radio-frequency electrical or magnetic stimuli are implemented for spin quantum-state control,[6,7] but the time limit of such operations are at best nanoseconds. More lately, femtosecond or picosecond optical pulses have enabled ultrafast spin manipulation at exceptionally high speeds.[8-15]

In spite of the success of various manipulation methods, from a viewpoint of practical applications, there still exist many shortcomings associated with current spin-host materials. Epitaxial QDs are fabricated using expensive, high-temperature and high-vacuum apparatus. Besides, a fundamental issue is that inter-level scattering and coupling to phonon bath can strongly damp the spin coherence. Consequently, spin manipulation of these QDs was typically accomplished at cryogenic temperatures of a few Kelvin.[9,15] By contrast, the defect or dopant spins in solids are highly isolated and can be manipulated at room temperature.[16] However, scaled-up production of these "point defects" might eventually become a challenge. For the scalable and sustainable implementation of spin-based QIS, it is desirable to develop low-cost materials whose spins can be coherently manipulated under ambient conditions.

The colloidal counterparts of QDs (also called nanocrystals) can be synthesized in large quantity in solution at low cost, yet with high precision in terms of size and shape control, and are particularly diverse for self-assembly or device integration.[17] However, spin manipulation for prototypical CdSe-based colloidal QDs has not been realized yet at room temperature.[10] We turn our focus to recently-developed lead halide perovskite (LHP) colloidal QDs.[18] Their spin-orbit coupling and electronic structure [19] have proved to be ideal for efficient spin injection by optical means [20-25], and their strong light-matter interaction should also facilitate spin manipulation based on an optical Stark effect (OSE) [26-28]. The challenge of spin manipulation in LHP QDs, however, is their rapid spin relaxation at room temperature (a few picoseconds),[24,25] likely limited by enhanced electron-hole exchange in these confined systems [29].

Here we combine interfacial charge-transfer chemistry of LHP QDs and femtosecond laser pulses to initialize, manipulate and readout single-hole spins at room temperature. We functionalize the surfaces of $CsPbBr_3$ QDs with anthraquinone (AQ) molecules. Upon preparation of a spin-polarized exciton in a QD using a circularly-polarized photon, the AQ can extract the electron on a sub-picosecond timescale, thus quenching the spin relaxation induced by electron-hole exchange. This results in long-lived hole spin precession about an applied transversal magnetic field until 100s of picoseconds, during which a second off-resonance laser pulse coherently rotates the hole spin around a longitudinal axis through the OSE. Taken together, the precession and rotation accomplish nearly complete quantum-state control of single-hole spins at room temperature. Notably, these operations are conducted for ensemble samples, which is advantageous over single-dot manipulation in that it can tolerate the loss of a few spins. For epitaxial QDs, ensemble-level spin manipulation has been very challenging due to QD inhomogeneity and required the use of special "mode-locking" techniques.[11,30]

**Sample information and experimental design.** Fig. 1 illustrates the sample information, optical layout and principle of our experiment. All measurements were performed at room temperature. CsPbBr$_3$ QDs of controllable sizes were synthesized using a hot-injection method;[24] details can be found in Methods. Fig. 1a shows the absorption spectra of two QD samples (QD1 and QD2) dispersed in hexane. Their transmission electron microscope (TEM) images in Supplementary Fig. 1 reveal monodisperse, cube-shaped dots with average edge lengths of ~4.2 and 4.6 nm for QD1 and QD2, respectively. The uniform quantum confinement results in a series of exciton peaks identifiable at room temperature, with the lowest ones at 470 and 481 nm for QD1 and QD2, respectively (Fig. 1a). The large energy separation between exciton peaks helps to suppress their coupling to phonon bath and to sustain spin coherence at room temperature. A carboxylated derivative of anthraquinone (AQ), which is a well-known electron acceptor,[31,32] was anchored onto QD surface through the carboxyl group (see Methods). The enhanced absorption of QD1-AQ and QD2-AQ compared to bare QD1 and QD2 in the ultraviolet can be attributed to the AQ molecules (Supplementary Fig. 2). On the basis of absorption spectra and extinction coefficients, we determine more than 200 AQ molecules bounded to each QD, although this is likely slightly overestimated (see Methods).

An approximate energy-level diagram (Supplementary Fig. 2) between CsPbBr$_3$ QDs [33] and AQ [31] predicts that electron transfer (ET) from photoexcited QDs to AQ is the only allowed charge/energy transfer pathway in the system. ET results in nearly quantitative quenching of the emissions of both QD1 and QD2. Full characterization of the charge-transfer processes is provided in Supplementary Fig. 3; the most important conclusion is that ET occurs mostly within 300 fs and the resulting $QD^+$-$AQ^-$ charge-separated states live much longer than 10 ns. The extremely rapid ET is mainly due to large amounts of AQ acceptors available to each QD, statistically enhancing the ET rate by 100s-fold.[34] The remaining hole could have long-lived spin polarization as these halide perovskites feature an "inverted" band structure compared to traditional semiconductors, with weak spin-orbit coupling in their valence band.[35]

The band-edge optical selection rules in $CsPbBr_3$ QDs are depicted in Fig. 1b in a quasi-particle representation.[24,27] In a cubic-symmetry, the valence-band states ($|s = 1/2, m_s = \pm 1/2\rangle$) are coupled to the conduction-band spin-orbit split-off states ($|j = 1/2, m_j = \mp 1/2\rangle$) through circularly-polarized photons ($\sigma^-$ and $\sigma^+$, respectively). Anisotropic exchange interaction due to QD shape anisotropy or lattice distortion may diagonalize the circularly-polarized excitons into linearly-polarized ones.[35-37] But the spin-selective OSE that we will show below indicates the spin selection rules still largely hold for the samples studied here at least at room temperature. Moreover, our primary focus here is single-hole states, for which exchange-induced splitting is eliminated. For the same reason, random orientation of the QDs in an ensemble should not complicate data interpretation,[38,39] provided that the rotational motion of these nanostructures in solution takes much longer time (nanoseconds[40]) than spin relaxation studied here.

In the optical experiment (Fig. 1c; detailed in Methods), we used a femtosecond laser amplifier to pump an optical parametric amplifier (OPA) to generate a tunable pump pulse. Another art of the fundamental beam was frequency-doubled to generate the rotation ("tipping") pulse at 515 nm. The spectra of the pump pulses for QD1 and QD2 are plotted in Fig. 1a, which are in resonance with their respective lowest-exciton peaks, whereas the tipping pulse is below their optical gaps and hence functions as off-resonance rotation. The white-light-continuum probe was generated by focusing a relatively weak 515 nm beam onto a sapphire window. The pulse polarizations were controlled by polarizers and waveplates. These pulses were focused onto the solution sample in a quartz cuvette; a transversal magnetic field ($B_z$) was applied on the sample (Voigt geometry). The pump and/or tipping pulses were modulated by choppers and the resulting absorption changes ($\Delta A$) were recorded by the probe pulse. The advantage of recording $\Delta A$ here over Faraday rotation in previous studies is the immediate accessibility to broad-band spectral information under various modulation schemes, as illustrated below.

**Room-temperature hole spin precession.** We first investigated hole spin injection and precession with the pump pulse on and the tipping pulse off (Fig. 1d). The pump power was minimized in order to ensure that only single-exciton states are generated in the QDs, without further complications from multiexciton Auger recombination.[41] The circularly-polarized pump pulse triggers sub-ps ET and leaves behind a spin-polarized hole. In the presence of the transversal magnetic field, the hole is in a coherent superposition of eigenstates $|\uparrow\rangle$ and $|\downarrow\rangle$ quantized by the field: $(|\uparrow\rangle \pm |\downarrow\rangle)/\sqrt{2}$ (Fig. 2a), situated at the *x*-axis of the Bloch sphere whose *z*-axis is aligned with $B_z$ (Fig. 2b). Due to the field-induced Zeeman splitting ($E_z$) of |↑> and |↓>, the coherent state is rotated on the equator plane with the angular frequency of $\omega = E_z/\hbar$ (i.e., the Larmor precession frequency). This precession can be directly visualized by a circularly-polarized probe pulse; see Fig. 2c for QD1-AQ and Supplementary Fig. 4 for QD2-AQ. As anticipated, the time-dependent ΔA spectra detected with co- and counter-polarized pump/probe configurations have exactly anti-correlated phases (compare left and right panels of Fig. 2c). In each configuration, the bleach at 470 nm and the induced absorption at 484 nm are contributed by hole-induced state-filling and Coulombic effects, respectively, and they have identical kinetics. The kinetics plotted at 484 nm are shown in Fig. 2d (top) and those at 470 nm are in Supplementary Fig. 5. On the basis of the signal sizes measured by co- and counter-configurations ($S_{co}$ and $S_{counter}$), we calculate the hole spin initialization efficiency (fidelity) of $\Phi_h = |S_{co} - S_{counter}|/|S_{co} + S_{counter}|$ ~50% (Supplementary Fig. 6 and Supplementary Text 1).

Our spin initialization method can be viewed as an "active" initialization as the hole is generated by dissociating the spin-polarized exciton injected by a pump photon, i.e., it carries the spin directly imprinted by the photon. By contrast, spin initialization of charge-doped QDs extensively adopted in previous studies[11-15] is more like a "passive" method in which the pump photon eliminates one of the pre-doped spins by promoting them to trion states and leaves the other type of spins in a polarized state. Although both methods work well, our new method does not require pre-doped samples and its interpretation is relatively more straightforward. It resembles, but is greatly simplified than, the electric field-induced exciton ionization method reported for epitaxial QDs.[14,42]

Fig. 2d (bottom) presents the hole spin precession kinetics in QD1-AQ obtained by taking the difference between co- and counter-ΔA (to remove any common background signals), at 484 nm. The kinetics can be fitted with a damped cosinoidal function: $S(t) \propto e^{-t/T_2^*} Cos(\omega t + \varphi)$, where $T_2^*$ is the transversal dephasing time (44.3 ps), $\omega$ is the precession angular frequency (5.66° per ps), and $\varphi$ is the initial phase close to zero (−1.47°); see Supplementary Table 1. Using $B_z$ = 0.65 T, the Landé g-factor of the hole is derived to be $|g_h| = \hbar\omega/\mu_B B_z$ = 1.73, which is larger than the reported value for bulk-like $CsPbBr_3$ (ref[38,39]) because quantum confinement modifies the g-factor[43]. Instead, it agrees reasonably well with $|g_h|$ in large-bandgap halide perovskites[44].

The zero-field kinetics of QD1-AQ are also plotted in Fig. 2d for comparison, which shows a relaxation time of 40.7 ps. The similarity between this lifetime and $T_2^*$ indicates that the Δg-mechanism of spin dephasing is negligible at this field strength, which can be attributed to the ensemble-level uniformity of our $CsPbBr_3$ QDs. Also notably, the zero-field kinetics of neutral QDs (not functionalized with AQ) of such sizes had a spin relaxation/dephasing time of only ~1 ps (Supplementary Fig. 7 and Supplementary Table 2).[24] The marked contrast between functionalized and unfunctionalized QDs substantiates that electron-hole exchange is indeed the limiting factor for the exciton spin lifetime, and that by removing the electron long-lived hole spin coherence is attainable at room temperature. Further, the exchange interaction between the hole and AQ radical anion should have a minor impact, on the basis of the QD size-independent hole spin lifetime (Supplementary Fig. 8).

**Ultrafast rotation using OSE**. We then studied the spin-selective OSE as a tool for coherent spin manipulation, in an experimental configuration with the pump pulse and $B_z$ off and the tipping pulse on (Fig. 1d). This measures the OSE of neutral exciton states instead of single-hole states in the sample, but it serves a good starting point to illustrate coherent manipulation.[45,46] As shown in Fig. 3a, when the system is coherently driven by, for example, a $\sigma^+$ tipping photon, under a quasi-particle representation, the Floquet states hybridize with only $|-1/2\rangle_h$ and $|+1/2\rangle_e$ states but not the other two states coupled to $\sigma^-$. As a result, a blue-shift of the transition can be detected with a $\sigma^+$ photon but not a $\sigma^-$ photon.

Indeed, we observed a lobe-shaped ΔA spectrum for QD1 with co-circularly polarized pump-probe beams), which is almost absent when the beams have counter circular polarizations (Fig. 3b); see Supplementary Fig. 9 for representative spectra of QD2. The signal is present only during tipping-probe cross-correlation time ($\tau_c$ ~540 fs at $1/e^2$ intensity). These are consistent with spin-selective OSE. The lobe-shaped ΔA spectral intensity increases with the tipping-power $P_{tip}$ (Fig. 3c), which can be quantified as OSE-induced splitting ($\delta_{OSE}$) between the $|\sigma^+\rangle$ and $|\sigma^-\rangle$ excitons. The splitting grows approximately linearly with $P_{tip}$ and reaches 9.65 meV at 9.72 GW/cm$^2$ (Fig. 3d). From the linear slope, we derive transition dipoles ($\mu$) of 21 and 24 Debye for QD1 and QD2, respectively; see Supplementary Text 2 for the calculation details[27]. The large $\mu$ is indicative of intrinsically strong light-matter interaction for CsPbBr$_3$ QDs, in contrast to a previous study on CdSe-based colloidal QDs which relied on a resonant plasmon enhancement effect to achieve sizable $\delta_{OSE}$.[10]

The strong light-matter interaction of CsPbBr$_3$ QDs should facilitate coherent spin manipulation using the OSE. To illustrate this, we can interpret the OSE that lifts the degeneracy of $|\sigma^+\rangle$ and $|\sigma^-\rangle$ excitons as an effective pseudo-magnetic field ($B_{eff}$).[8-10,45,46] The direction of this field is along the tipping beam (x-axis; Fig. 3e). If a coherent exciton state of $(|\sigma^+\rangle + |\sigma^-\rangle)/\sqrt{2}$ is prepared, the OSE is able to rotate the exciton state in QD1 with an angle up to 275° (4.8 radian) around $B_{eff}$ (see Supplementary Fig. 10 and Supplementary Text 3). Note this estimation is made under the assumption of a linear increase of $B_{eff}$ with $\delta_{OSE}$ (and hence $P_{tip}$).[46] For a hole state, we estimate the rotation angle to be up to 137.5° (2.4 radian) by assuming that electron and hole states equally share $\delta_{OSE}$ of the exciton state.[8] An alternate interpretation of spin rotation using strong pulses is the stimulated Raman transition theory.[12,15,47]

**Room-temperature coherent hole spin manipulation.** With the tools of spin precession and manipulation at hand, we explored complete quantum-state control of single-hole spins in QD1-AQ with both the pump (chopped) and tipping (unchopped) pulses as well as $B_z$ on (Fig. 1d). The pump-tipping delay ($t_{tip}$) is controlled to investigate the tipping effects at various positions on the Bloch sphere (z-axis aligned to $B_z$). Fig. 4a (top) is the spin precession kinetics tipped at $t_{tip}$ = 17.2 ps, i.e., when the state is $(|\uparrow\rangle + i|\downarrow\rangle)/\sqrt{2}$ on the y-axis of the Bloch sphere, with a tipping power of 9.72 GW/cm$^2$. A large amplitude change and sign-switch from the untipped kinetics is achieved; see the corresponding Bloch-sphere representation in Fig. 4b. The difference of tipping with $\sigma^+$ or $\sigma^-$ pulses is negligible (Supplementary Fig. 11). By contrast, when the the tipping acts at $t_{tip}$ = 31.7 ps ($|\uparrow\rangle - |\downarrow\rangle$)/$\sqrt{2}$; *x*-axis) with the same power, a negligible amplitude change is observed (Fig. 4a; middle). Overall, tipping the states at *y*-axis and *x*-axis represents the most and least obvious spin manipulations, respectively. Fig. 4a (bottom) presents an intermediate case with the tipping acting at $\omega t + \varphi$ ~4.4 radian between *y*-axis and *x*-axis. Similar results for QD2 are presented in Supplementary Fig. 12. All the tipped kinetics can be well fitted using Bloch-sphere analysis; see Supplementary Text 4 and Supplementary Tables 3-6 for details.

We examined the tipping-power ($P_{tip}$) dependence with $t_{tip}$ fixed at 17.2 ps (Fig. 4c), i.e., tipping at the *y*-axis. The tipping angle is calculated as $\theta_{tip}$ = arccos($A_t/A_u$), with $A_t$ and $A_u$ being the tipped and untipped signal amplitudes, respectively.[8,10] As presented in Fig. 4d, $\theta_{tip}$ of QD1 increases sublinearly with $P_{tip}$ ($\theta_{tip} \propto P_{tip}^{0.636}$) until it reaches ~$2\pi/3$ radian at ~10 GW/cm$^2$. The maximum $\theta_{tip}$ was solely limited by the laser power in our setup as no sign of sample damage was observed at the largest $P_{tip}$. For QD2 with a larger transition dipole and a smaller tipping detuning, $\theta_{tip}$ increases with $P_{tip}$ also sublinearly ($\propto P_{tip}^{0.634}$), but more steeply than QD1, reaching $0.56\pi$ radian at 3.58 GW/cm$^2$. A further increase of $P_{tip}$, however, results in real excitation of trion states because the absorption onset of QD2 is closer to the tipping pulse than QD1 (Fig. 1a). We expect that $\pi$-radian tipping should be achievable for both QD1 and QD2 by tailoring the tipping photon energy and bandwidth, as well as by increasing the tipping power.

Notably, the sublinear scaling of $\theta_{tip}$ with $P_{tip}$ observed herein contradicts the simple assumption of $B_{eff}$ scaling linearly with $P_{tip}$ made previously[46], but it is similar to a previous study on epitaxial QDs[15]. As explained in ref[15], when the Rabi energy ($\hbar\Omega_R$) of the interaction between the electric field of the tipping pulse and the QD transition dipole becomes comparable to the detuning energy ($\Delta$), the so-called adiabatic elimination approximation breaks down, and the excitation of virtual population has to be considered. For example, at $P_{tip}$ of 9.72 GW/cm$^2$, $\hbar\Omega_R$ has reached 67.3 meV, which is indeed comparable with $\Delta$ ~230 meV in our experiment. A four-level master-equation simulation in ref[15] produces $\theta_{tip} \propto P_{tip}^{0.65}$, which is strikingly close to our experimental results.

The complete set of initialization, manipulation and readout of single-hole spins in an ensemble of CsPbBr$_3$ perovskite QDs achieved herein at room temperature is a very promising result. It establishes the feasibility of quantum information processing using low-cost, solution-grown samples under ambient conditions. Further prolonging the room-temperature spin coherence time to nanoseconds, thus enabling $10^4$-$10^5$ operations using femtosecond pulses, is the next step in the roadmap.

**Methods**

**Chemicals.** Cesium carbonate (Cs$_2$CO$_3$, Sigma-Aldrich, 99.9%), lead(II) bromide (PbBr$_2$, Alfa Aesar, 98%), zinc bromide (ZnBr$_2$, Alfa Aesar, 99.9%), oleic acid (OA,

Sigma-Aldrich, 90%), oleylamine (OAm, Acros Organics, 80–90%), 1-octadecene (ODE, Sigma-Aldrich, 90%), methyl acetate (Energy Chemical, 99% Extra Dry), anthraquinone-2-carboxylic Acid (AQ, Alfa Aesar, 98%). All chemicals were used directly without any further purification.

**Synthesis of CsPbBr$_3$ QDs.** CsPbBr$_3$ QDs were synthesized following previously reported procedures[24,48]. The synthesis started with preparation of Cs-oleate precursors. 0.25g Cs$_2$CO$_3$, 0.8 g OA, and 7 g ODE were loaded into a 50 mL 3-neck flask and vacuum-dried for 1 h at 120 °C using a Schlenk line. The mixture was heated under N$_2$ atmosphere to 150 °C until all the Cs$_2$CO$_3$ powders were dissolved. The Cs-oleate precursor solution was kept at 100 °C to prevent precipitation of Cs-oleate out of ODE. In another 50 mL 3-neck flask, the precursor solution of Pb and Br was prepared by dissolving 225 mg PbBr$_2$ and 552 mg ZnBr$_2$ in a mixture of ODE (18 mL), OA (9 mL), and OAm (9 mL). After the precursor solution of Pb and Br were vacuum-dried for 1 h at 120 °C and the temperature reached 140 °C under N$_2$ atmosphere, 1.2 mL of Cs precursor solution was injected to initiate the reaction. The reaction was quenched after 30 s by cooling the flask in an ice bath. After the crude solution was cooled down to room temperature, the product was centrifuged at 3500 rpm for 20 min to remove the unreacted salts as the precipitate, and the perovskite QDs dispersed in the supernatant were collected. Further, the QDs were precipitated by adding methyl acetate dropwise until the mixture just turned turbid to avoid decomposition of the QDs. After being centrifuged at 6000 rpm for 5 min, the precipitate was dried and dissolved in hexane. Adding extra methyl acetate to the second supernatant could precipitate smaller QDs. This process could be repeated to harvest target-size QDs.

**Preparation of QDs-AQ complexes.** The $CsPbBr_3$ QD-AQ complexes were prepared by adding anthraquinone-2-carboxylic Acid (AQ) powders into the $CsPbBr_3$ QDs solution followed by stirring for 30 min and filtration. On the basis of the reported extinction coefficients of QDs[49] and the one measured for AQ (~5500 $M^{-1}cm^{-1}$ at 325 nm), there are, on average, ~280 and 220 AQ molecules for each QD1 and QD2, respectively. The solubility of AQ molecules in hexane is negligible, but in QD-hexane solution, which contains excessive amounts of OA and OAm ligands, AQ becomes slightly soluble. Thus, the molecular numbers above are slightly overestimated. Nevertheless, previous nuclear magnetic resonance measurements for similar $CsPbBr_3$ QD-molecule systems[50] suggest that the majority of acceptor molecules were indeed bounded to QD surfaces.

**Femtosecond transient absorption.** Femtosecond transient absorption were carried out using a Pharos femtosecond laser system (Light Conversion; 1030 nm, FWHM 230 fs, 20 W) and Orpheus-HP optical parameter amplifier (Light Conversion). Repetition frequency of the Pharos femtosecond laser system is tunable from 1kHz to 100 kHz and was set at 10 kHz for current experiments. The 1030 nm output laser was split into two beams with 80/20 ratio. The 80% part was used to pump an Orpheus-HP optical parameter amplifier (OPA; Light Conversion) to generate a wavelength tunable pump beam. The remaining 1030 nm beam from the OPA was focused into a 2 mm thick BBO crystal to generate a 515 nm tipping beam. A notch filter with a center wavelength at 514±2 nm and FWHM of 17 nm was used to remove 1030 nm photons from the tipping pulses. The 20% part was further split into two parts with 75/25 ratio. The 75% part was attenuated with a neutral density filter and focused into a BBO crystal to generate a 515 nm beam, which was further focused into a Sapphire crystal to generate a white light continuum used as the probe beam. The probe beam was focused with an Al parabolic reflector onto the sample. After the sample, the probe beam was collimated and then focused into a fiber-coupled spectrometer with CMOS line scan camera and detected at a frequency of 10 kHz.

The intensity of the pump and tipping pulses used in the experiment was controlled by a variable neutral-density filter wheel. The delay between the pump and probe pulses was controlled by a motorized delay line and the delay between the pump and tipping pulses was controlled by a homemade delay line. The pump or tipping beam was chopped by a synchronized chopper at 5 kHz and the absorbance change was calculated with adjacent probe pulses. The probe, pump and tipping beams were focused and spatially overlapped on the sample with a spot size of 240 μm, 270 μm and 355 μm, respectively (at $1/e^2$ intensity). The spot size was measured using the knife-edge method. The pulse durations were ~540 fs at $1/e^2$ intensity. Circular polarizations of the pump, tipping and probe beams were controlled by quarter waveplates. The magnetic field direction was perpendicular to the laser beams, which was provided by an electromagnet (EM3; Beijing Jinzhengmao Technology Co.).

**Acknowledgments**

K.W. acknowledged financial support from the the Ministry of Science and Technology of China (2018YFA0208703), the Chinese Academy of Sciences (YSBR-007), and Dalian Institute of Chemical Physics (DICP I201914).

**Author contributions**

K.W. conceived the idea and designed the project. X.L. and Y.H. synthesized the samples, measured the spectroscopy with the help of J.Z., and analyzed the data. K.W. wrote the manuscript with inputs from all authors. X.L. and Y.H. contributed equally.

**Competing interests:** Authors declare no competing interests.

**Data and code availability:** All data and code are available in the main text or the supplementary materials and can be obtained upon reasonable request from K.W. (kwu@dicp.ac.cn).

**Supplementary Materials:** Supplementary Figures 1-12**;** Supplementary Tables 1-6**;** Supplementary Texts 1-4.

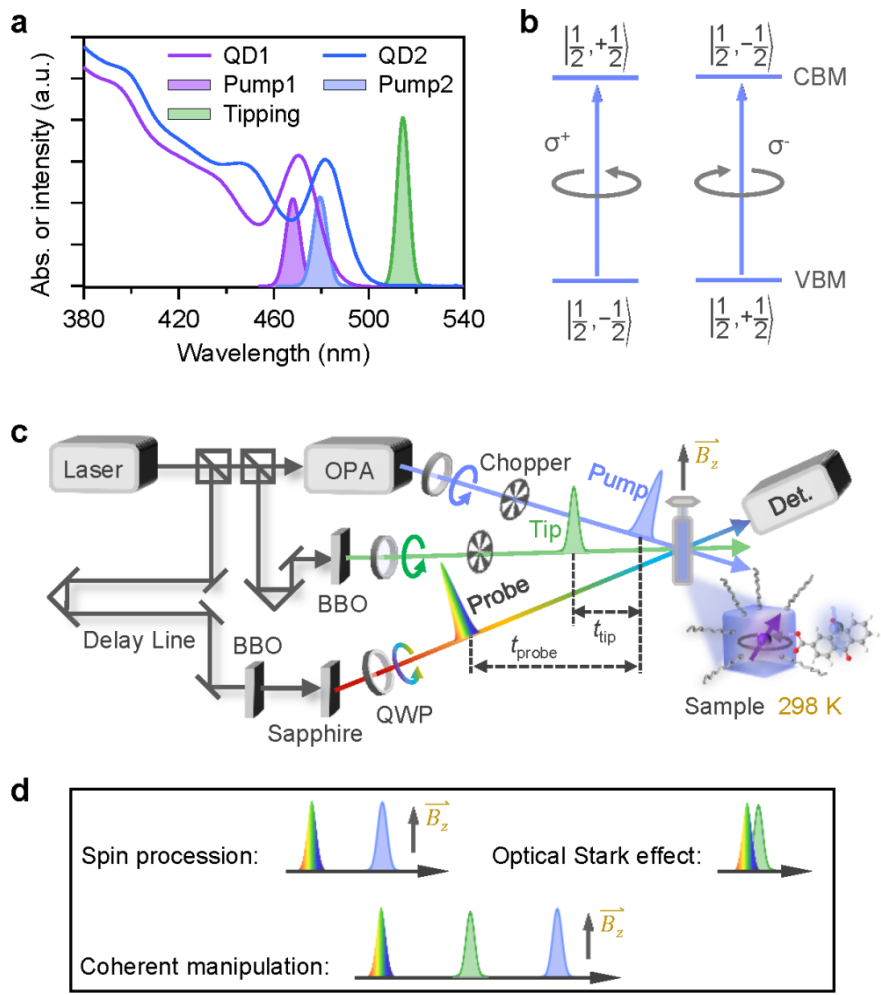

**Figure 1. System design and experimental setup.** (a) Absorption spectra of CsPbBr$_3$ QD1(purple solid line) and QD2 (blue solid line). The spectra of the pump and the tipping pulses (shaded pulses) are included for comparison. (b) Band-edge optical selection rules in CsPbBr$_3$ QDs in a quasi-particle representation. Circularly polarized σ$^+$ and σ$^-$ photons are selectively coupled to different spin-transitions. (c) Optical layout of the spin manipulation experiments. OPA: optical parametric amplifier; BBO: Barium borate crystal; QWP: quarter waveplate. (d) The different experimental schemes of the pulse sequences to study spin precession, optical Stark effect and coherent spin manipulation.

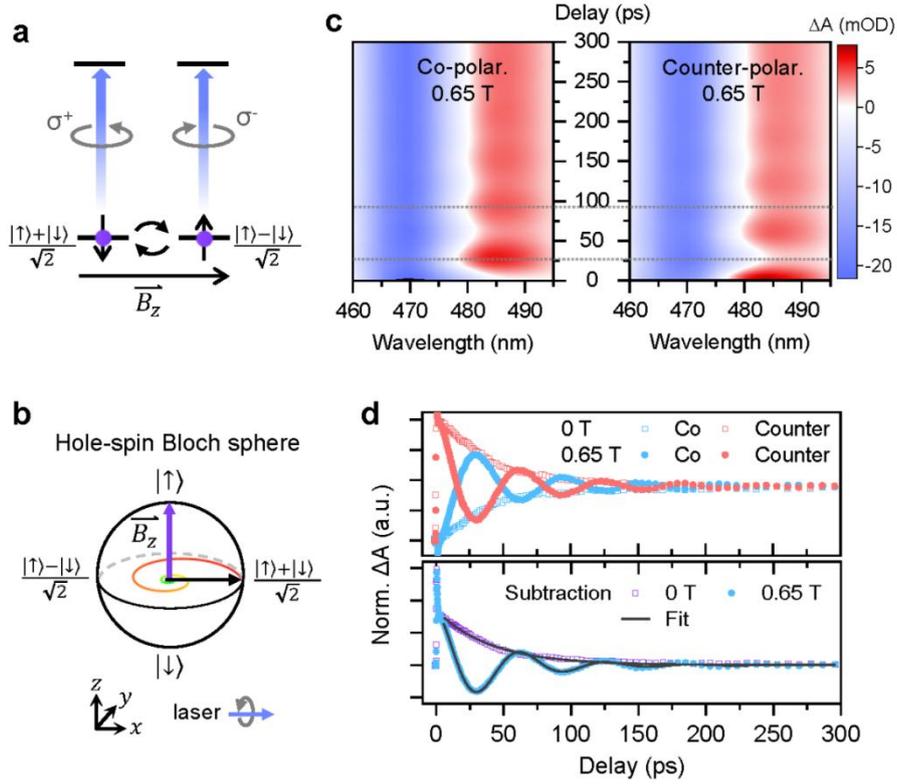

**Figure 2. Room-temperature hole spin precession in CsPbBr$_3$ QDs.** (a) Energy level diagram in the presence of a transversal magnetic field ($B_z$). After removing the conduction band electron using an acceptor, the valance band hole oscillates between ($|\uparrow\rangle \pm |\downarrow\rangle)/\sqrt{2}$, where $|\uparrow\rangle$ and $|\downarrow\rangle$ are eigenstates quantized by the field. (b) Bloch sphere representation of hole spin precession. The *z*-axis is aligned with $B_z$. (c) Two-dimensional pseudo-color transient absorption (TA) spectra of QD1-AQ measured with (left) co- ($\sigma^+/\sigma^+$) and (right) counter-polarized ($\sigma^-/\sigma^+$) pump/probe beams at $B_z$ = 0.65 T. (d) Top, TA kinetics probed at 484 nm revealing opposite phases measured with co- (blue filled circles) and counter-polarized (red filled circles) pump/probe beams at 0.65 T; Bottom, extracted spin precession kinetics at 484 nm (blue filled circles) through subtraction between the two curves above. The open squares are the corresponding kinetics measured at 0 T for comparison. The black solid lines are fits.

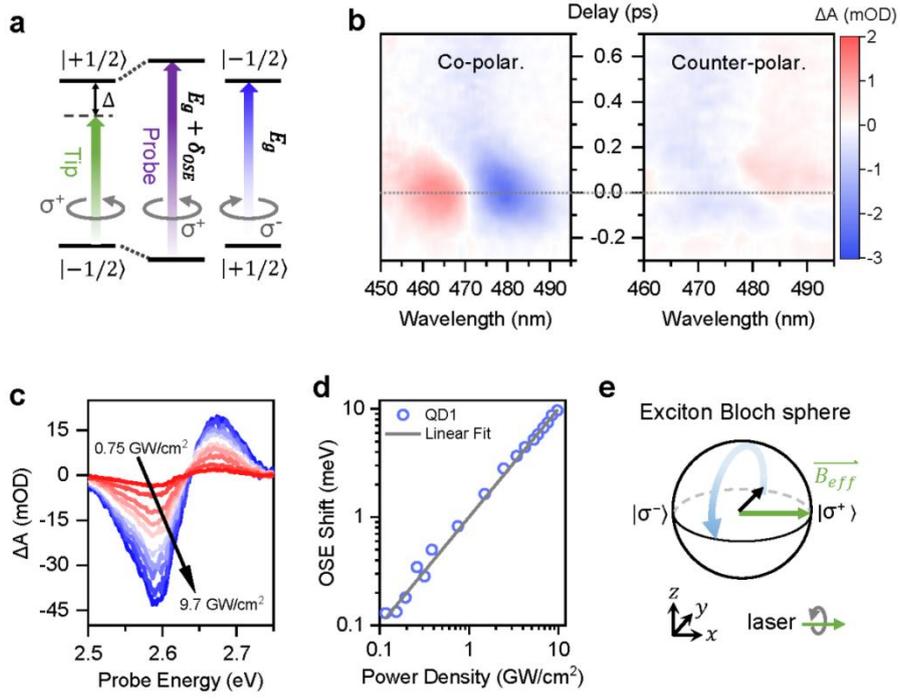

**Figure 3. Ultrafast rotation using optical Stark effect (OSE)**. (a) Scheme of the spin-selective OSE in CsPbBr$_3$ QDs. $\Delta$ is the detuning of the driving photon ($\sigma^+$) than the optical transition energy ($E_g$), and $\delta_{OSE}$ is the OSE-induced blue-shift of the transition energy. For simplicity, only the Floquet state of "dress-up" from $|-1/2\rangle_h$ is shown, while the other one associated with "dress-down" from $|+1/2\rangle_e$ is omitted. (b) Two-dimensional pseudo-color TA spectra of QD1-AQ measured with (left) co- ($\sigma^+/\sigma^+$) and (right) counter-polarized ($\sigma^-/\sigma^+$) tipping/probe beams with a tipping power density of 0.38 GW/cm$^2$. (c) The time-zero OSE spectra at varying tipping power densities. (d) The OSE shift ($\delta_{OSE}$) as a function of tipping power density (blue circles) and a linear fit (gray solid line). (e) Bloch sphere representation of coherent exciton state rotation by the OSE-induced effective magnetic field ($B_{eff}$) along the *x*-axis.

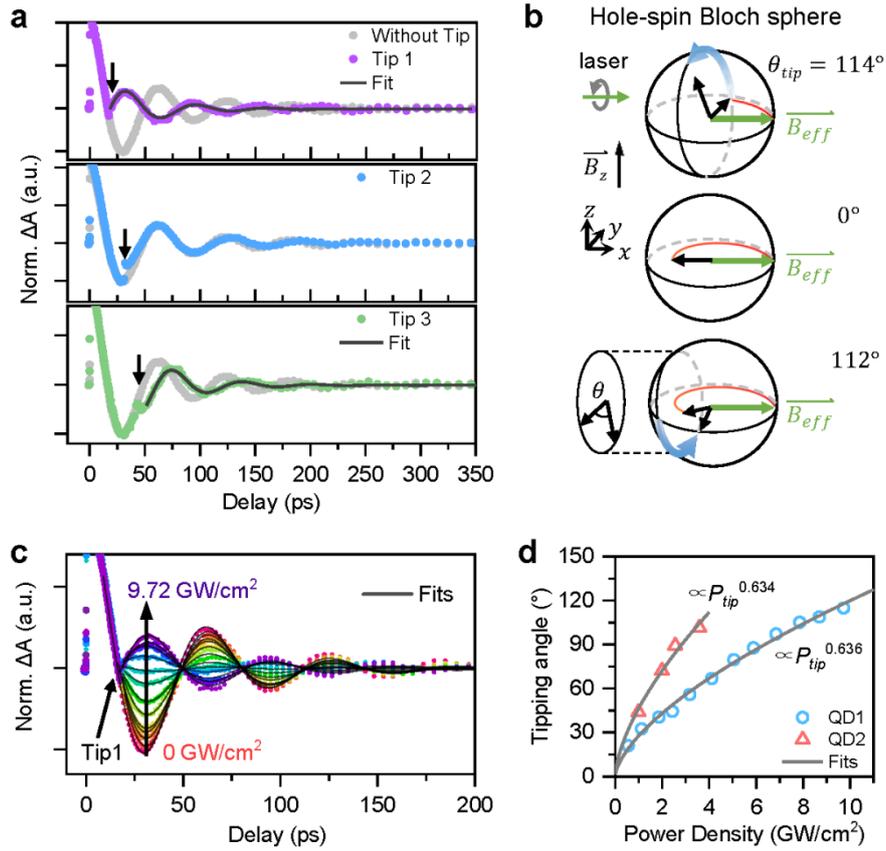

**Figure 4. Room-temperature hole spin manipulation in CsPbBr$_3$ QDs.** (a) The untipped spin precession kinetics (gray circles) and the tipped kinetics (colored circles) with the tipping pulse acting at different times (top: 17.2 ps; middle, 31.7 ps; bottom; 42.0 ps) indicated by black arrows. The tipping power density is 9.72 GW/cm$^2$. The solid lines are fits. (b) Corresponding Bloch sphere representation of coherent hole spin manipulation using the $B_{eff}$ of the tipping pulse. (c) Tipping power density dependent spin precession kinetics with the tipping time fixed at 17.2 ps (colored dots). The solid lines are their fits. (d) Tipping angle as a function of tipping power density for QD1 (blue circles) and QD2 (red triangles). The gray solid lines are their fits.

Supplementary Materials for

# Room-temperature coherent optical manipulation of single-hole spins in solution-grown perovskite quantum dots

Xuyang Lin[1,2†], Yaoyao Han[1,2†], Jingyi Zhu[1], Kaifeng Wu[1,2]*

Correspondence to: kwu@dicp.ac.cn

**This PDF file includes:**

Supplementary Texts 1-4

Supplementary Figures 1 to 12

Supplementary Tables 1 to 6

**Supplementary Text**

1. Estimation of spin injection efficiency

As shown in Supplementary Fig. 6A, at near time-zero, the TA spectra of CsPbBr$_3$ QD1-AQ acquired with circularly polarized pump/probe configurations contain an exciton bleach (XB) feature at ~470 nm due to the state-filling effect and a photoinduced absorption on the red side (PIA2) at ~484 nm due to biexcitonic interaction. This interaction can be attributed to an attraction between $\sigma^+$ and $\sigma^-$ excitons; therefore it is strong for counter-polarizations but almost negligible for co-polarizations [1-3]. In addition to PIA2, there is another PIA on the blue side at ~450 nm (PIA1). The nature of this feature remains unclear. It has been assigned as the "exciton-activated forbidden transition"[4] in strongly-confined perovskite QDs. Alternatively, it can be associated with the blueshift of band-edge excitons or the redshift of higher-energy excitons.

In general, the exciton spin injection efficiency can be calculated using:[5]

$$\Phi_X(t = 0\ ps) = |\frac{S_{co} - S_{counter}}{S_{co} - S_{counter}}| \tag{S1},$$

where $S_{co}$ and $S_{counter}$ are the integral areas of TA spectra measured under co- and counter-polarizations, respectively. If we take integrals for the whole spectral window, with the underlying assumption that PIA1 arises from the blueshift of band-edge excitons, we obtain unrealistic value of $\Phi_X(t = 0\ ps) \sim 200\%$. This contradiction indicates that PIA1 cannot be assigned this way.

If we make alternate assumptions that PIA1 arises from an activated forbidden transition or from the redshift of higher-energy excitons, i.e., it is not associated with band-edge excitons, we should exclude PIA1 from the integrals and integrate only XB and PIA2. In doing so, we obtain $\Phi_X(t = 0\ ps) = 70\%$ (Supplementary Fig. 6A). *The calculated imperfect $\Phi_X$ has multiple possible reasons: i) Direct exclusion of PIA1 leads to underestimation of $\Phi_X$, as PIA1 might spectrally overlap with XB; ii) The degree of circular polarization is not perfect, typically on the level of ~95%; iii) The instrument resolution (~540 fs at $1/e^2$ intensity) is insufficient to resolve all the spin relaxation processes, i.e., there might be very fast exciton spin relaxation not captured by our TA, and thus our $\Phi_X(t = 0\ ps)$ is already a partially relaxed value.*

If electron extraction from QD1 by surface-tethered AQ molecules completely overwhelms exciton spin relaxation, the hole spin injection efficiency $\Phi_h$ should track $\Phi_X$. However, electron transfer from QDs to molecules often has very heterogeneous kinetics.[6] While there are fast components on the femtosecond timescale being much faster than the ~1 ps exciton spin relaxation, there are also slower components on the order of a few ps. The net result is a lower $\Phi_h$ compared to $\Phi_X$. In the case for QD1-AQ, we calculate it from the spectral integrals at $t = 2.7$ ps, when electron transfer and exciton spin relaxation finish but hole spin relaxation has not started yet. We obtain $\Phi_h(t = 2.7\ ps) = 51\%$.

The calculations were also made for QD2-AQ, and we obtain $\Phi_X(t = 0\ ps) = 67\%$ and $\Phi_h(t = 2.7\ ps) = 47\%$ for this sample; see Supplementary Fig. 6 panels C and D, respectively.

## 2. Estimation of OSE shift, Rabi energy and transition dipole

For an arbitrary absorption lineshape $A(E)$ with the absorption peak at $E_0$, the OSE energy shift ($\delta_{OSE}$) can be estimated by defining a spectral weight transfer (SWT) :[7,8]

$$\text{SWT} = \int_0^{E_0} \delta A(E) dE \qquad (S2),$$

where $\delta A(E)$ is the spectral shift. It is easy to show that with the defined SWT, $\delta_{OSE}$ due to the OSE can be calculated as:

$$\delta_{OSE} = -\frac{\text{SWT}}{A(E_0)} \qquad (S3).$$

The calculation is illustrated for QD1 in Fig. 3 and QD2 in Supplementary Fig. 9.

$\delta_{OSE}$ is related to the Rabi energy ($\hbar\Omega$) and pump energy ($\Delta$) as:

$$\delta_{OSE} = \sqrt{(\hbar\Omega)^2 + \Delta^2} - \Delta \approx \frac{(\hbar\Omega)^2}{2\Delta} \propto \text{Power density} \qquad (S4).$$

The approximation is valid for the case of $\Delta \gg \hbar\Omega$. For QD1, using the $\delta_{OSE}$ of 9.65 meV at tipping power density of 9.72 GW/cm² with $\Delta$ of 230 meV, we obtain $\hbar\Omega = 67.3$ meV or 66.4 meV using the exact or approximate relations, respectively, in eq S4.

Using the definition of $\hbar\Omega$, eq. S4 can be rewritten as:[9]

$$\delta_{OSE} \approx \frac{(\hbar\Omega)^2}{2\Delta} = \frac{\mu^2 \langle E^2 \rangle}{\Delta} \qquad (S5),$$

where $\mu$ is the QD transition dipole and $\langle E^2 \rangle$ is the time-averaged value of the electric field squared. For QDs dispersed in a solvent, $\langle E^2 \rangle$ inside QDs can be related to the pump intensity $I_0$ via:

$$\langle E^2 \rangle = \frac{|F^2| I_0}{\epsilon_0 \sqrt{\epsilon_s} c} \qquad (S6),$$

where $\epsilon_S$ is the dielectric constant of the solvent, $\epsilon_0$ the vacuum permittivity, $c$ the light velocity and $F$ the local field factor accounting for the difference between the electric field inside the QD and that in the solvent. $F$ has been derived as:

$$F = \frac{3\epsilon_S}{\epsilon_{QD} + 2\epsilon_S} \quad (S7),$$

where $\epsilon_{QD}$ is the high frequency dielectric constant of the QD.

We adopt $\epsilon_{QD} = 4.8$ [10] for CsPbBr$_3$ QDs in the calculation. Combining eqs S5-S7, we obtain $\mu = 21$ Debye for QD1 and 24 Debye for QD2. Uncertainties associated with the measurement of beam spot sizes and pulse durations can introduce errors to the calculations of $I_0$ and thus the final values of $\mu$, but the large values of $\mu$ are in general consistent with previous studies[9].

### 3. Estimation of effective pseudomagnetic field and spin rotation induced by OSE

The splitting of the exciton spin-states $|\pm 1>$ for CsPbBr$_3$ QDs in a pseudomagnetic field $B_{eff}$ induced by the OSE can be expressed as:

$$\delta_{OSE} = g_X \mu_B B \quad (S8),$$

where $\delta_{OSE}$ is the OSE shift, $g_X$ is the exciton g-factor and $\mu_B = 9.27 \times 10^{-24} J \cdot T^{-1}$ is the Bohr magneton. Using $g_X \sim 2.4$ reported in refs [11,12], the $B_{eff}$ is estimated as ~69 T for QD1 with $\delta_{OSE} = 9.65$ meV under a tipping power intensity of 9.72 GW/cm$^2$, or ~46 T for QD2 $\delta_{OSE} = 6.37$ meV under a tipping power intensity of 3.58 GW/cm$^2$. Note that in ref [13], the relation of $\delta_{OSE} = 2g_X \mu_B B$ was used, with the factor of 2 different from eq. S8. This simply depends on how the g-factor of excitons is defined. We use eq. S8 because $g_X \sim 2.4$ reported in refs [11,12] was derived from Zeeman splitting being equal to $g_X \mu_B B$.

Supplementary Fig. 10 illustrates the process to calculate the expected spin rotation angle ($\theta_{OSE}$) by the pseudomagnetic field $B_{eff}$. In a quasi-particle representation, we assume valence and conduction-band contribute equally to the blue shift of the band gap (exciton), i.e. $\delta_{VB} = \delta_{CB} = 1/2 \delta_{OSE}$ (Supplementary Fig. 10A). The hole-spin $\theta_{OSE}$ can be calculated using:[14]

$$\theta_{OSE} = \frac{1}{\hbar} \int_{-\infty}^{\infty} \delta_{VB}(t)\, dt \quad (S9),$$

where $\hbar$ is the reduced Plank constant.

The time-dependent $\delta_{OSE}(t)$ was estimated through the kinetics of OSE signals (Supplementary Fig. 10B, duration ~540 fs at 1/e$^2$), by scaling the maximum of the latter to the

$\delta_{OSE}$ determined with the SWT method above (Supplementary Fig. 10C). The hole-spin $\theta_{OSE}$ calculated using eq. S9 was plotted in Supplementary Fig. 10D for QD1 and QD2. According to this estimation, for QD1 at a tipping power density of 9.72 GW/cm², the estimated rotation angle for the hole-spin should reach 2.4 radian. Note, however, that this estimation only serves as a general guide, not only because there are uncertainties associated with the calculation of $\delta_{OSE}$, but also the adiabatic approximation implicit to this model might break down at high tipping power densities [15].

4. Analyzing the TA kinetics of coherent spin manipulation

The spin procession kinetics obtained by taking the difference between co- and counter-$\Delta A$ can be fitted with a damped oscillatory function:

$$\Delta A(t) = A_u e^{-\frac{t}{T_2^*}} Cos(\omega t + \varphi_u) \qquad (S10),$$

where $A_u$ is the (untipped) signal amplitude, $T_2^*$ is the transversal dephasing time, $\omega$ is the procession angular frequency, and $\varphi$ is the initial phase. The fitting parameters are listed in Supplementary Table 1 for QD1-AQ and QD2-AQ samples. This damped oscillation can be visualized on the Bloch sphere drawn in Fig. 2b. The spin vector initialized along the $x$-axis makes Larmor precession about the $z$-axis defined by the external field $B_z$. The damping (dephasing) is manifested as the gradual shortening of the vector.

The TA kinetics obtained with the tipping pulse on was separated into two parts. One is before the tipping pulse, identical to the untipped kinetics, and the other is after tipping. The latter part is also fitted with a damped oscillatory function:

$$\Delta A(t) = A_t\, e^{-\frac{t}{T_2^*}} Cos(\omega t + \varphi_t) \qquad (S11),$$

where $T_2^*$ and $\omega$ are identical to the untipped parameters in eq. S10, but $A_t$ and $\varphi_t$ can be altered by the tipping pulse. The parameters are tabulated in Supplementary Tables 3-6.

The principles can be illustrated by the Bloch spheres in Fig. 4b. A special case is that $B_{eff}$ (along the $x$ direction) is perpendicular to the spin vector, for which we have negligible $\Delta\varphi$, whereas $\theta_{tip} = \Delta\theta = arcCos(\frac{A_t}{A_u})$. For arbitrary angles between $B_{eff}$ and the spin vector, we still have $\Delta\theta = arcCos(\frac{A_t}{A_u})$ but a nonzero $\Delta\varphi = \varphi_t - \varphi_u$. Using the pump-tipping delay ($t_{tip}$) and the fitting parameters of eq. S11, we can obtain the spin vector's coordinates before and after the rotation on the Bloch sphere. We project these spin vectors back to the *y-z* plane and calculate the tipping angle $\theta_{tip}$.

Note that occasionally we find the long-lived population signal (i.e., at >200 ps) after spin dephasing to be slightly different for tipped and untipped kinetics. This difference might be caused by excitation of the hole by the tipping pulse to certain surface/trap states, thereby reducing the overall population of the holes inside the QDs. The difference becomes obvious at high tipping power densities. To account for this difference, the kinetic traces after the tipping pulses were scaled by a factor of 1-1.3 such as to match their long-lived tails to the untipped traces, with the scaling factors summarized in Supplementary Table 3.

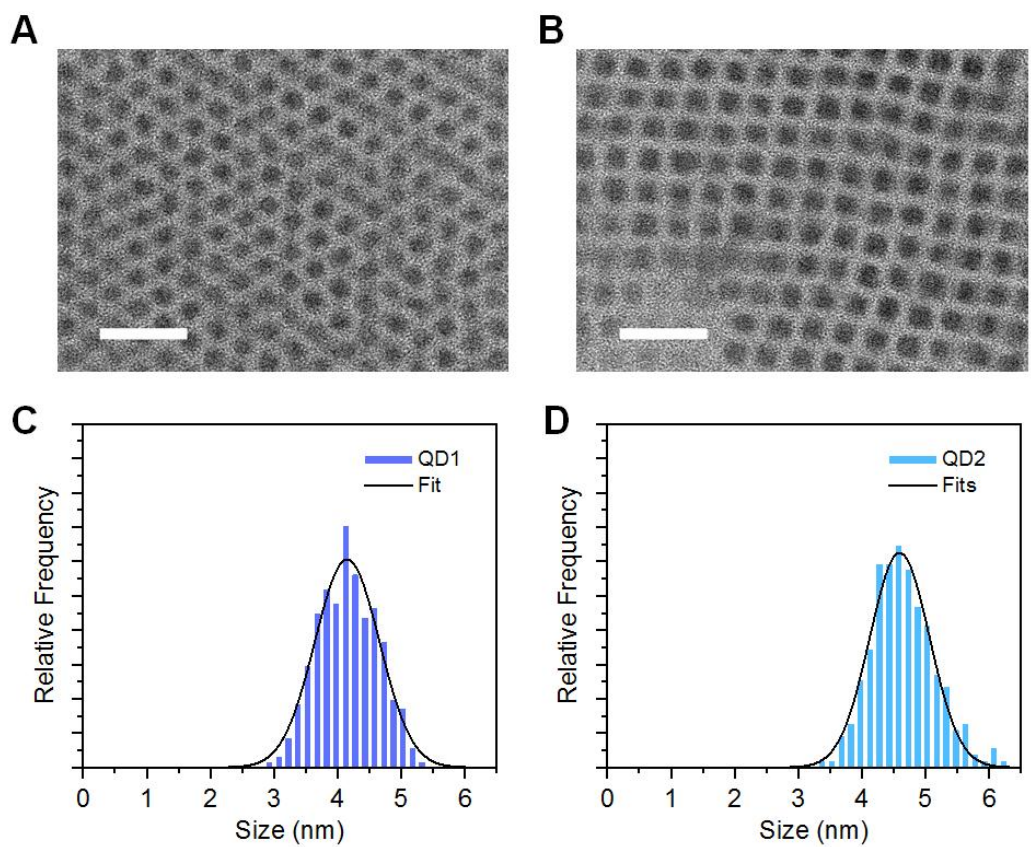

**Supplementary Figure 1. Transmission electron microscope (TEM) images of CsPbBr$_3$ QDs and their size distributions.**

(A, B) TEM images of (A) QD1 and (B) QD2. Both scale bars are 20 nm. (C, D) size distributions of (C) QD1 with an average edge length of 4.2 nm and (D) QD2 with an average edge length of 4.6 nm.

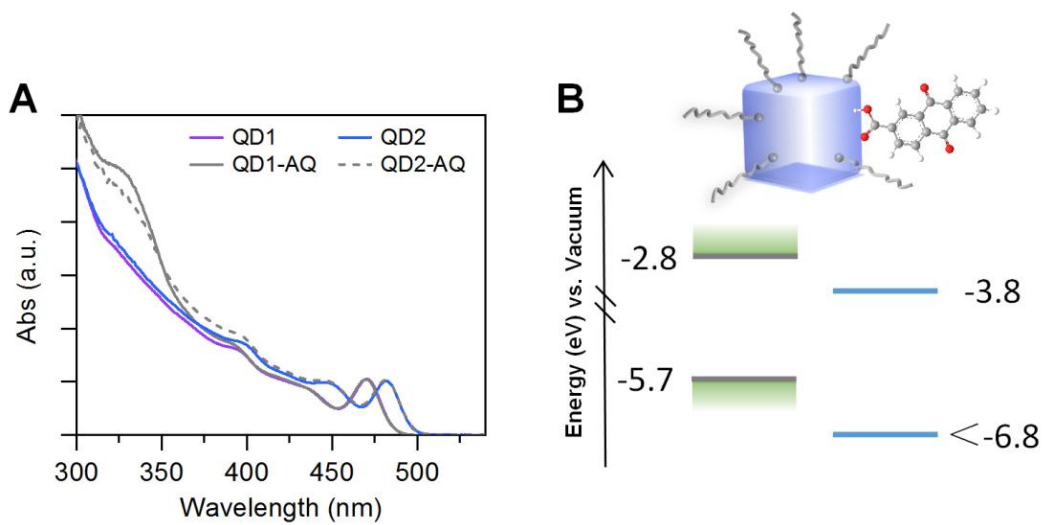

**Supplementary Figure 2. UV-Vis absorption spectra and schematic energy levels of CsPbBr$_3$ QDs and QD-AQ complexes.**

(A) UV-vis absorption spectra of QD1 (purple solid line), QD2 (blue solid line), QD1-AQ (gray solid line) and QD2-AQ (gray dash line). The difference between QDs and QD-AQ complexes below 390 nm can be attributed the absorption of AQ molecules. (B) Schematic energy level alignment of QDs reported in ref. [16] and AQ reported in ref. [17].

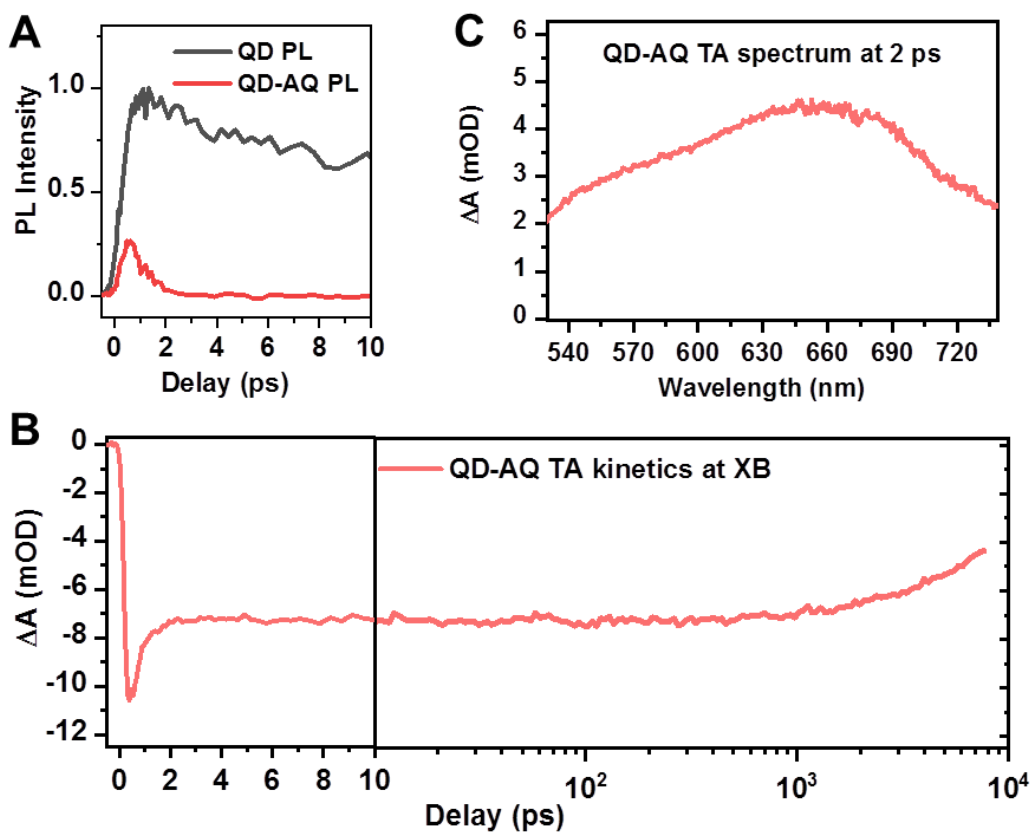

**Supplementary Figure 3. Femtosecond electron transfer from CsPbBr$_3$ QDs to AQs.**

(A) Ultrafast PL dynamics probed by femtosecond upconversion technique for free QDs (gray solid line) and QD-AQ complexes (red solid line). The two samples were measured under identical conditions (the same sample absorbance at 400 nm pump wavelength and the same pump power and collection conditions). But the QD-AQ sample has an instantaneously reduced PL signal amplitude, suggesting that electron transfer occurs in <<300 fs (IRF). The observed decay in 2 ps corresponds to slow components of the heterogeneous electron transfer dynamics. (B) TA dynamics probed at the QD exciton bleach center for the QD-AQ sample. The rapid decay within the first 2 ps is consistent with ultrafast PL measurements. After this rapid decay, the remaining bleach signal are extremely long-lived, which can be assigned to state-filling and Coulomb signals induced by the hole remaining in the QD. The charge-separated state (QD$^+$-AQ$^-$) shows only minor decay within 10 ns. (C) The reduced AQ radical anion (AQ$^-$) feature probed in the QD-AQ sample at 2 ps.

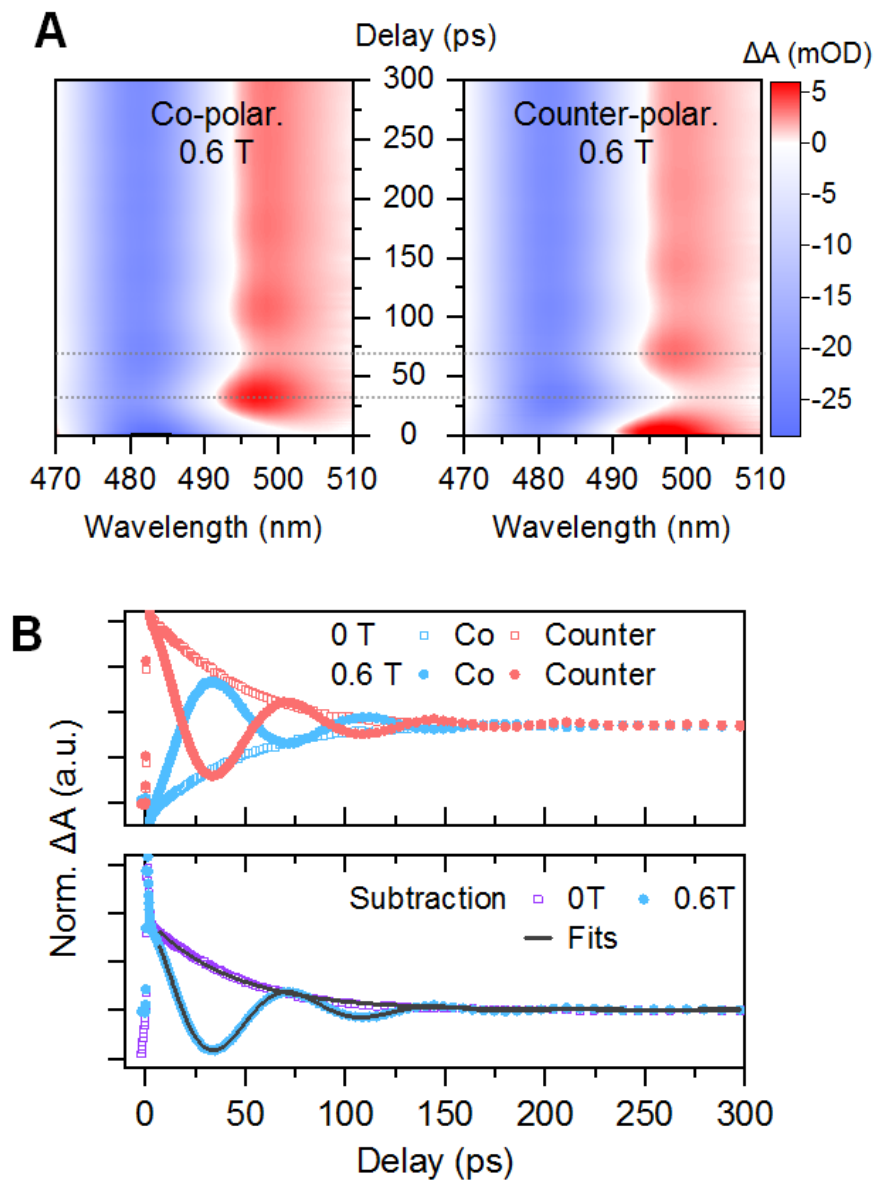

**Supplementary Figure 4. Room-temperature hole spin precession in CsPbBr$_3$ QD2-AQ.**

(A) Two-dimensional pseudocolor transient absorption (TA) spectra of QD2-AQ measured with (left) co- ($\sigma^+/\sigma^+$) and (right) counter-polarized ($\sigma^-/\sigma^+$) pump/probe beams at $B_z = 0.6$ T. (B) Top, TA kinetics probed at 500 nm revealing opposite phases measured with co- (blue filled circles) and counter-polarized (red filled circles) pump/probe beams at 0.6 T; Bottom, extracted spin precession kinetics at 500 nm (blue filled circles) through subtraction between the two curves above. The open squares are the corresponding kinetics measured at 0 T for comparison. The black solid lines are fits.

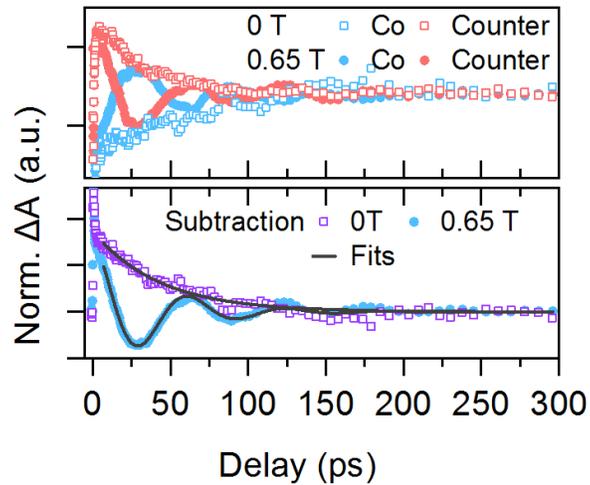

**Supplementary Figure 5. Room-temperature hole spin precession in CsPbBr$_3$ QD1-AQ probed at 470 nm.**

The top panel reveals opposite phases measured with co- (blue filled circles) and counter-polarized (red filled circles) pump/probe beams at 0.65 T. The bottom panel shows extracted spin precession kinetics at 470 nm (blue filled circles) through subtraction between the two curves above. The open squares are the corresponding kinetics measured at 0 T for comparison. The black solid lines are fits. The results are identical to those probed at 484 nm.

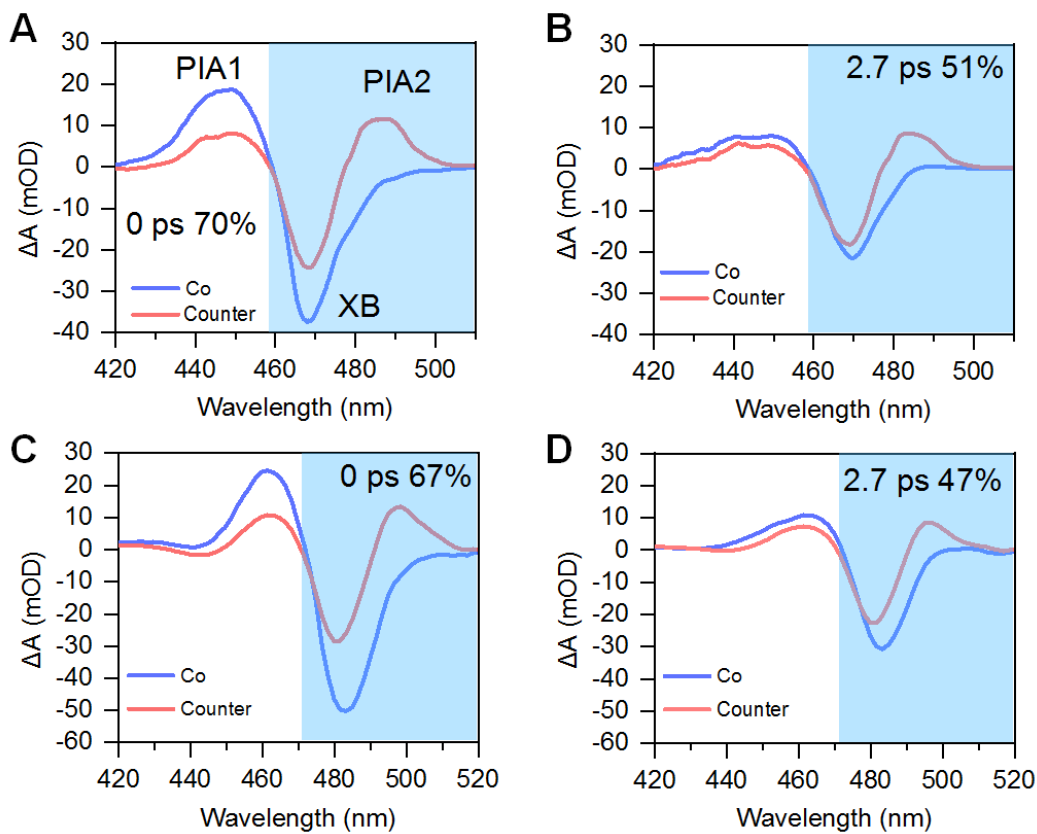

**Supplementary Figure 6. Calculation of spin injection efficiency.**

(A) Time-zero and (B) 2.7 ps co (blue line) and counter (red line) TA spectra of QD1-AQ complexes. The blue shaded regions are used to integrate the spectral areas. The exciton spin injection efficiency calculated at time-zero is 70%, whereas the hole spin injection efficiency calculated at 2.7 ps is 51%. (C,D) Similar plots for QD2-AQ complex.

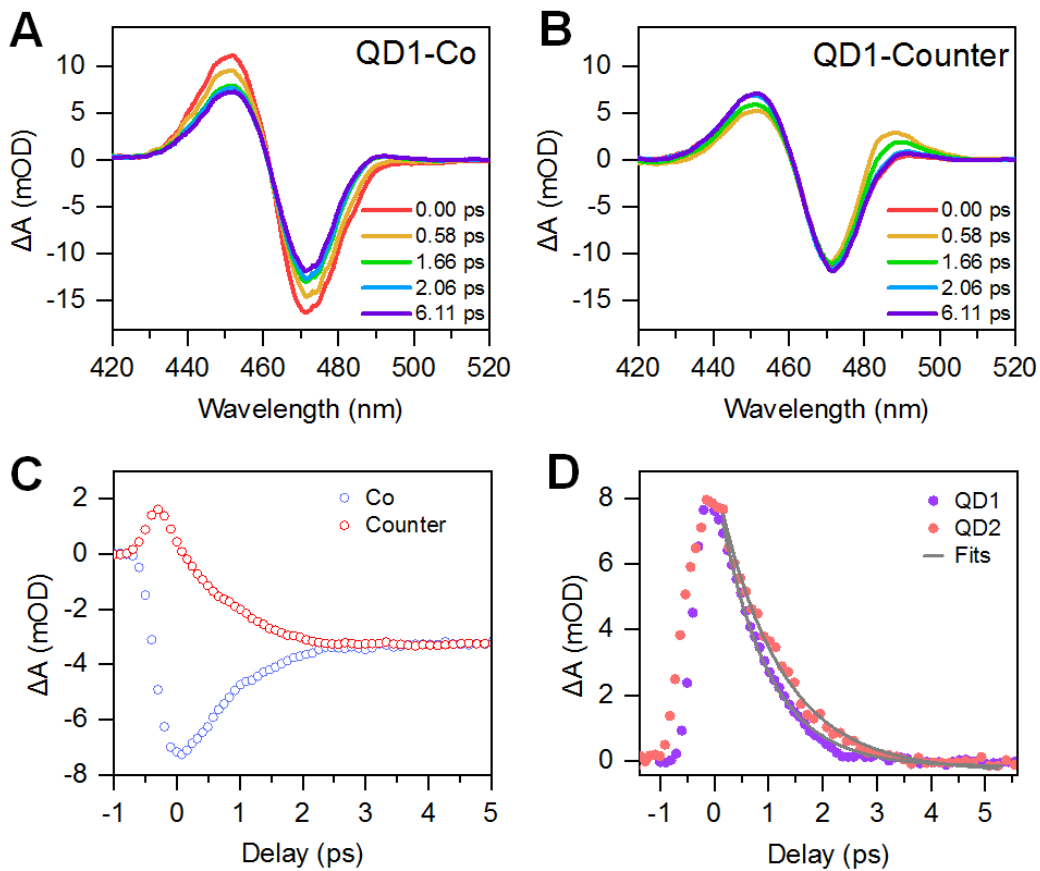

**Supplementary Figure 7. Exciton spin relaxation in CsPbBr$_3$ QDs.**

(A) Co and (B) counter circularly polarized TA spectra of QD1 at selected delays following 470 nm excitation. (C) Co (blue open circle) and counter (red open circle) TA kinetics probed at 482 nm. (D) Exciton spin relaxation kinetics, obtained by subtracting co- and counter-kinetics, for QD1 (purple circle) and QD2 (red circle). Gray lines are their single-exponential fits with time constants of 0.88 ps and 1.15 ps, respectively.

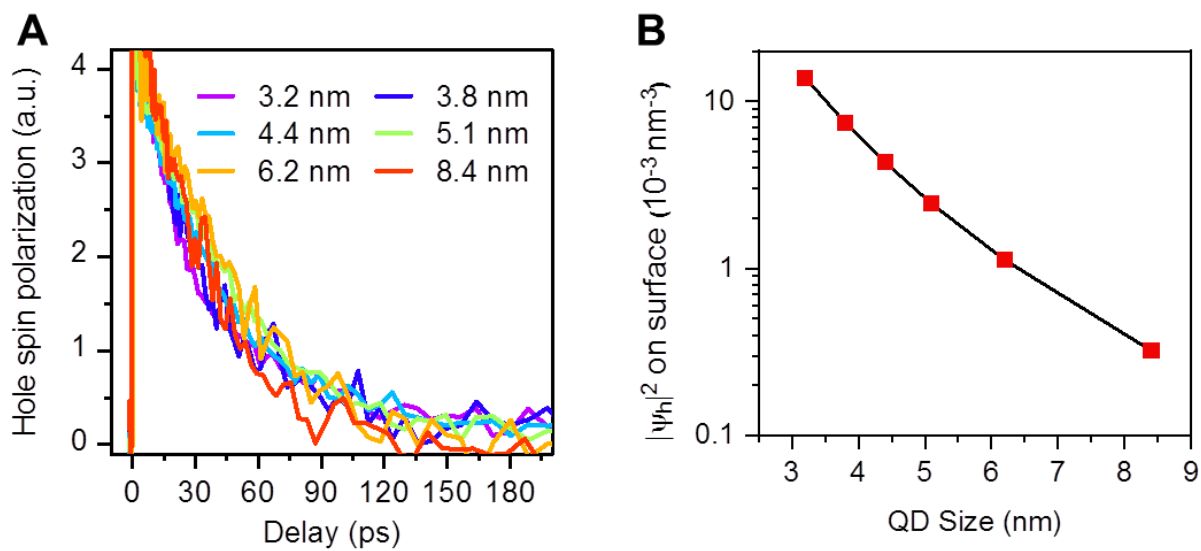

**Supplementary Figure 8. Insensitivity of hole spin lifetime to QD size.**

(A) Hole spin polarization decay kinetics measured for a series of $CsPbBr_3$ QD-AQ samples with QD edge lengths varying from 3.2 to 8.4 nm, which are almost identical within the noise level. (B) Hole wavefunction squared (i.e., probability density) at the QD surface as a function QD size, calculated using the method detailed in ref[18]. According to previous studies, exchange-type of interactions (e.g., charge transfer[17], triplet energy transfer[18]) depends sensitively on the QD wavefunction tunneled to the surface. For this reason, we argue that if the hole remaining in $CsPbBr_3$ QD has a strong exchange interaction with the reduced $AQ^-$ species, the hole spin polarization decay time should also depend sensitively on QD size, which is in direct contrast to the observation in (a).

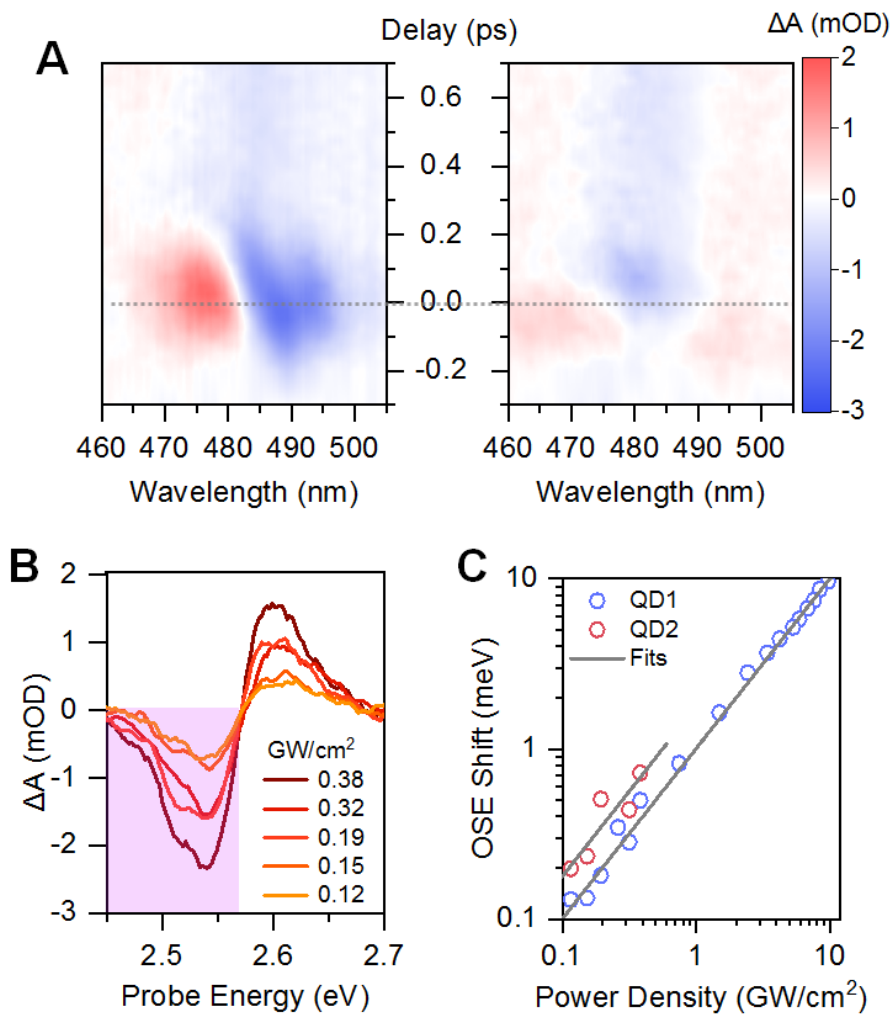

**Supplementary Figure 9. Optical Stark effect (OSE) of QD2-AQ.**

(A) Two-dimensional pseudocolor TA spectra of QD2-AQ measured with (left) co- (σ+/σ+) and (right) counter-polarized (σ-/σ+) tipping/probe beams with a tipping power density of 0.38 GW/cm². (B) The time-zero OSE spectra at varying tipping power densities. The shaded regions are used to calculate the spectral weight transfer (SWT). (C) The OSE shift ($\delta_{OSE}$) as a function of tipping power density for QD1-AQ (Blue circles) and QD2-AQ (red circles) and their linear fits (gray solid lines).

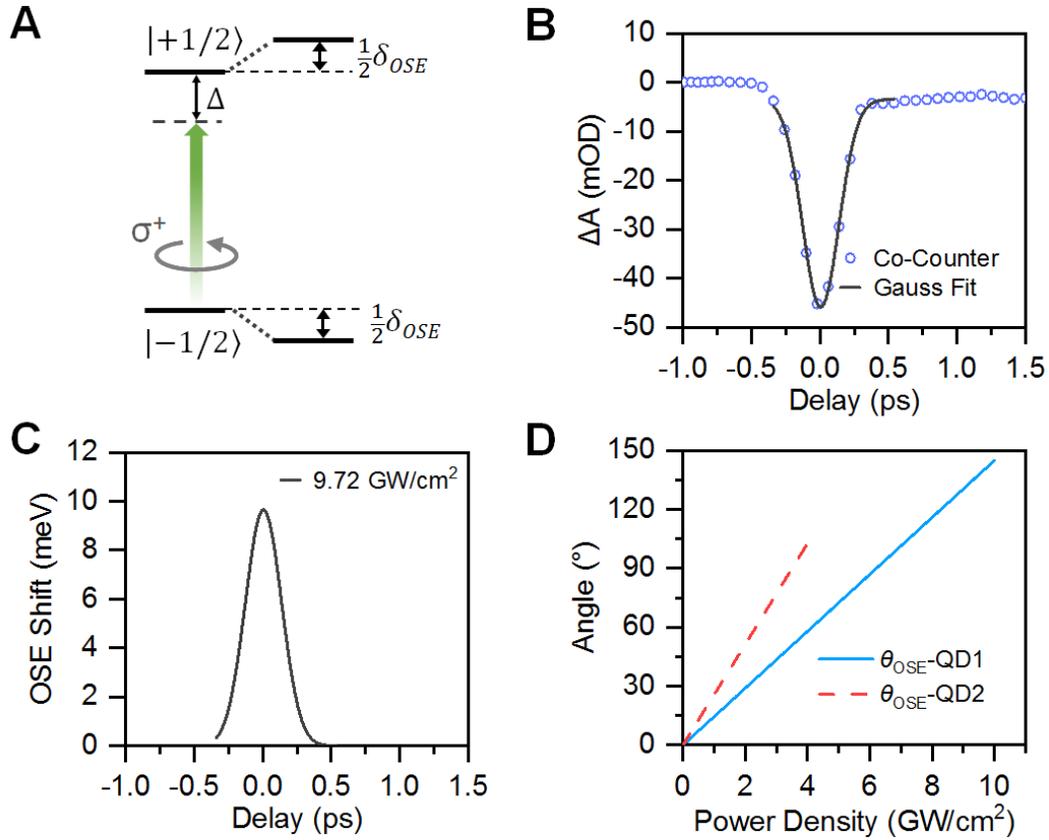

**Supplementary Figure 10. Estimation of spin rotation angles by Stark shifts.**

(A) Scheme of the spin-selective OSE in CsPbBr$_3$ QDs. $\delta_{OSE}$ is the OSE-induced blueshift of the transition energy. The shift of VBM and CBM is assumed to be $\delta_{OSE}/2$. (B) Subtraction of co- and counter-TA kinetics probed at 479 nm for QD1-AQ system with a tipping power density of 9.72 GW/cm$^2$ (blue open circles). Gray line is its Gaussian fit. (C) OSE shift as a function of time with a tipping power density of 9.72 GW/cm$^2$. (D) Estimated hole-spin rotation angles ($\theta_{OSE}$) for QD1 (blue solid) and QD2 (red dashed) as a function of tipping power density.

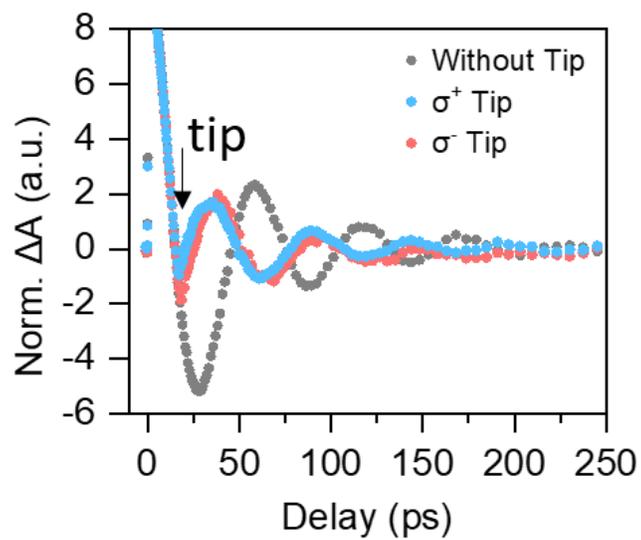

**Supplementary Figure 11. Influence of the tipping pulse helicity.**

Kinetics at 484 nm for QD1-AQ system using $\sigma^+$ (blue circle) and $\sigma^-$ tipping pulses under 9.72 GW/cm$^2$ and 0.65 T. The difference is negligible, except that the tipping times of $\sigma^+$ and $\sigma^-$ are slightly different.

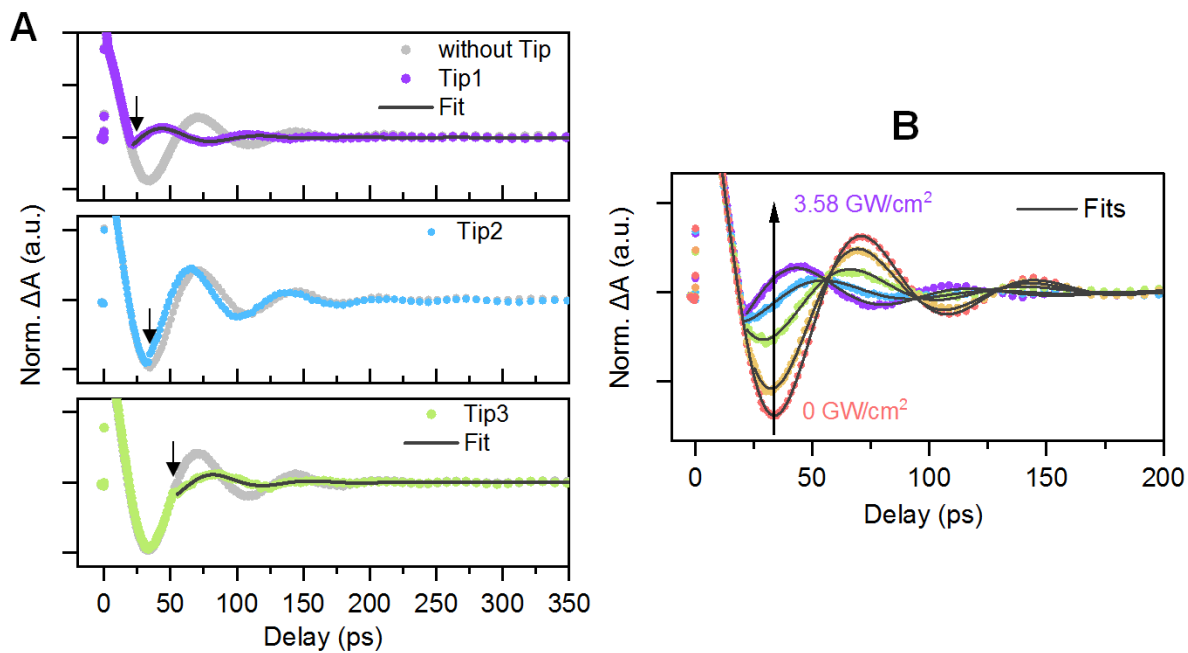

**Supplementary Figure 12. Room-temperature hole spin manipulation in QD2-AQ.**

(A) The untipped spin precession kinetics (gray circles) and the tipped kinetics (colored circles) with the tipping pulse acting at different times (top: 20.0 ps; middle, 34.1 ps; bottom; 52.0 ps) indicated by black arrows. The tipping power density is 3.58 GW/cm$^2$. The solid lines are fits. (B) Tipping power density dependent spin precession kinetics with the tipping time fixed at 20.0 ps (colored dots). The solid lines are their fits.

**Supplementary Table 1.**

Parameters of damped oscillatory fits of untipped TA kinetics.

|  | $A_u$ | $T_2^*$ (ps) | $\omega$ (°/ps) | $\varphi_u$(°) |
|---|---|---|---|---|
| QD1-AQ (0.65 T) | 0.01044 | 44.31 | 5.66 | -1.47 |
| QD2-AQ (0.60 T) | 0.00933 | 44.42 | 4.88 | 0.14 |

**Supplementary Table 2.**

Zero-field exciton/hole relaxation time of QD1, QD1-AQ, QD2 and QD2-AQ.

|  | QD1 | QD2 | QD1-AQ | QD2-AQ |
|---|---|---|---|---|
| relaxation time (ps) | 0.88 | 1.15 | 40.66 | 41.91 |

**Supplementary Table 3.**

Parameters of damped oscillatory fits of tipped TA kinetics for QD1-AQ under varying power densities at 17.2 ps pump-tip delay.

| Power density (GW/cm$^2$) | $A_t$ | $\varphi_t$(°) | $\theta_{tip}$(°) * | Correction factor |
|---|---|---|---|---|
| 0 | 0.01044 | -1.47 | / | / |
| 0.56 | 0.00975 | -7.62 | 20.94 | 1 |
| 1.12 | 0.00881 | -7.83 | 32.45 | 1 |
| 1.87 | 0.00793 | -7.78 | 40.57 | 1 |
| 2.43 | 0.00747 | -9.11 | 44.31 | 1 |
| 3.18 | 0.00586 | -10.56 | 55.85 | 1 |
| 4.11 | 0.00412 | -12.59 | 66.76 | 1 |
| 5.05 | 0.00188 | -16.47 | 79.63 | 1 |
| 5.89 | 0.000414 | -41.51 | 87.73 | 1 |
| 6.85 | -0.0013 | -0.44 | 97.15 | 1.158 |
| 7.85 | -0.00273 | -6.93 | 105.16 | 1.206 |
| 8.70 | -0.00427 | -13.11 | 108.77 | 1.247 |
| 9.72 | -0.00243 | -13.47 | 114.47 | 1.32 |

* $\theta_{tip} = arcCos(\frac{A_t}{A_u})$. Note $A_t$ switches its sign as the power density exceeds 5.89 GW/cm$^2$

**Supplementary Table 4.**
Parameters of damped oscillatory fits of tipped TA kinetics for QD1-AQ at 42.0 ps pump-tip delay.

| Delay3 | $A_t$ | $\varphi_t(°)$ | $\Delta\theta(°)$ | $\Delta\varphi(°)$ | $\theta_{tip}(°)$ * |
|---|---|---|---|---|---|
| untipped | 0.01022 | -1.50 | / | / | / |
| tipped | 0.00856 | -72.2 | 32.1 | -72.2 | 112 |

* $\theta_{tip}$ is calculated from spin vector analysis on the Bloch sphere.

**Supplementary Table 5.**
Parameters of damped oscillatory fits of tipped TA kinetics for QD2-AQ under varying power densities at 20.0 ps pump-tip delay.

| Power density (GW/cm$^2$) | $A_t$ | $\varphi_t(°)$ | $\Delta\theta(°)$ | $\Delta\varphi(°)$ | $\theta_{tip}(°)$ * |
|---|---|---|---|---|---|
| 0 | 0.00933 | 0.14 | / | / | / |
| 0.994 | 0.00681 | 6.63 | 43.1 | 6.6 | 43.97 |
| 1.99 | 0.00329 | 23.2 | 69.4 | 23.2 | 72.06 |
| 2.56 | 0.00152 | 78.6 | 80.6 | 78.6 | 89.36 |
| 3.58 | 0.00234 | 134 | 75.5 | 134 | 101.47 |

* $\theta_{tip}$ is calculated from spin vector analysis on the Bloch sphere.

**Supplementary Table 6.**
Parameters of damped oscillatory fits of tipped TA kinetics for QD2-AQ at 52.0 ps pump-tip delay.

| Delay3 | $A_t$ | $\varphi_t(°)$ | $\Delta\theta(°)$ | $\Delta\varphi(°)$ | $\theta_{tip}(°)$ * |
|---|---|---|---|---|---|
| untipped | 0.01068 | 0.91 | / | / | / |
| tipped | 0.00358 | -54.8 | 70.4 | -54.8 | 83.4 |

* $\theta_{tip}$ is calculated from spin vector analysis on the Bloch sphere.